\documentclass[fleqn,usenatbib]{mnras}
\usepackage{newtxtext,newtxmath}
\usepackage[T1]{fontenc}
\DeclareRobustCommand{\VAN}[3]{#2}
\let\VANthebibliography\thebibliography
\def\thebibliography{\DeclareRobustCommand{\VAN}[3]{##3}\VANthebibliography}

\usepackage{amsmath}
\usepackage{graphicx}
\usepackage{enumerate}   
\usepackage{color,xcolor}
\hypersetup{colorlinks=true,citecolor=blue,urlcolor=blue,linkcolor=blue}

\def\gr{$\gamma$-ray}

\title[Magnetized gal. outflows contribution to FRM]{The contribution of magnetized galactic outflows to extragalactic Faraday rotation}

\author[A. Arámburo-García et al.]{Andrés Arámburo-García$^{1}$\thanks{aramburo@lorentz.leidenuniv.nl},
Kyrylo Bondarenko$^{2,3,4}$\thanks{kyrylo.bondarenko@sissa.it},
Alexey Boyarsky$^{1}$\thanks{boyarsky@lorentz.leidenuniv.nl},
Andrii Neronov$^{5,6}$\thanks{andrii.neronov@apc.in2p3.fr},
\newauthor
Anna Scaife$^{7}$\thanks{anna.scaife@manchester.ac.uk} and Anastasia Sokolenko$^{8,9}$\thanks{ sokolenko@kicp.uchicago.edu}
\\
$^{1}$Institute Lorentz, Leiden University, Niels Bohrweg 2, Leiden, NL-2333 CA, the Netherlands\\
$^{2}$IFPU, Institute for Fundamental Physics of the Universe, via Beirut 2, I-34014 Trieste, Italy\\
$^{3}$SISSA, via Bonomea 265, I-34132 Trieste, Italy\\
$^{4}$INFN, Sezione di Trieste, SISSA, Via Bonomea 265, 34136, Trieste, Italy\\
$^{5}$Universit\'e  de  Paris  Cite,  CNRS,  Astroparticule  et  Cosmologie, F-75013  Paris,  France\\
$^{6}$Laboratory  of  Astrophysics,  Ecole  Polytechnique  Federale  de  Lausanne,  CH-1015,  Lausanne,  Switzerland\\
$^{7}$Department of Physics \& Astronomy, University of Manchester, UK\\
$^{8}$Theoretical Astrophysics Department, Fermi National Accelerator Laboratory, Batavia, Illinois, 60510, USA\\
$^{9}$Kavli Institute for Cosmological Physics, The University of Chicago, Chicago, IL 60637, USA
}

\date{Accepted XXX. Received YYY; in original form ZZZ}

\pubyear{2022}

\begin{document}
\label{firstpage}
\pagerange{\pageref{firstpage}--\pageref{lastpage}}
\maketitle

\begin{abstract}
Galactic outflows driven by star formation and active galactic nuclei blow bubbles into their local environments, causing galactic magnetic fields to be carried into intergalactic space.
We explore the redshift-dependent effect of these magnetized bubbles on the Faraday Rotation Measure (RM) of extragalactic radio sources.
Using the IllustrisTNG cosmological simulations, we separate the contribution from magnetic bubbles from that of the volume-filling magnetic component expected to be due to the seed field originating in the Early Universe. We use this separation to extract the redshift dependence of each component and to compare TNG model predictions with observation measurements of the NRAO VLA Sky Survey (NVSS).
We find that magnetized bubbles provide a sizeable contribution to the extragalactic RM, with redshift-independent $\langle |{\rm RM}| \rangle \simeq 13$~rad/m$^2$ for sources at redshifts $z\ge 2$. This is close to the mean residual RM of $16$~rad/m$^2$ found from NVSS data in this redshift range.
Using the IllustrisTNG simulations, we also evaluate a simple model for the contribution to residual RM from individual host galaxies and show that this contribution is negligible at high-redshift. 
While the contribution from magnetic bubbles in the IllustrisTNG model is currently compatible with observational measurements of residual RM, the next-generation RM sky surveys, which will be free from the wrapping uncertainty, have larger statistics and better sensitivity should be able to observe predicted flat contribution from magnetic bubbles at large redshifts. This should allow to experimentally probe magnetic bubbles and check models of galaxy feedback in cosmological simulations.
\end{abstract}


\section{Introduction}

The Faraday rotation technique provides a powerful probe of astrophysical magnetic fields across different elements of Large Scale Structure (LSS), from galaxies \citep{2015A&ARv..24....4B} to galaxy clusters and the inter-cluster medium \citep{galaxies6040142}. It has also been used to constrain the magnetic field strength in the intergalactic medium \citep{1994RPPh...57..325K,Blasi:1999hu,Neronov:2013lta,Pshirkov:2015tua,Aramburo-Garcia:2022ywn}. Measurements of the weakest intergalactic magnetic fields (IGMF) using the Faraday rotation technique are challenging. The observational signal is determined by the  Rotation Measure (RM), which is an integral along the line of sight toward the source of the polarized signal:
\begin{equation} 
    \text{RM} = \frac{e^3}{2\pi m_{\rm e}^2}\int \frac{n_{\rm e} B_{\parallel}}{(1+z)^2} \frac{{\rm d}\ell}{{\rm d}z} {\rm d}z,
    \label{eq:FRMeq}
\end{equation}
where $e,\,m_{\rm e}$ are the charge and mass of the electron,  $n_{\rm e}$ is the density of free electrons in the medium, $z$ is the redshift, and $B_{||}$ is the magnetic field component parallel to the line of sight.  
This integral has contributions from the intergalactic medium (IGM) and the Milky Way along a line-of-sight (LoS). The Milky Way part of LoS has a small length but large $n_{\rm e}$ and $B_{||}$ values, while the IGM part is significantly longer but has smaller $n_{\rm e}$ and $B_{||}$.
In addition to the contributions from the Galactic RM and the IGM, the integral in Eq.~\eqref{eq:FRMeq} also has a contribution from the source host galaxy and possibly from parts of the LoS passing through magnetized regions of other galaxies occasionally found close to the LoS \citep{Bernet:2008qp}. Uncertainties in modeling the Galactic component of the RM \citep{jansson12,2012A&A...542A..93O,Oppermann:2014cua,Hutschenreuter2020}, of the source host galaxy, as well as the elements of LSS along the LoS, limit the sensitivity of searches for the contribution from the intergalactic medium and IGMF in the integral.

Detailed modeling for both the evolution of the primordial field and the baryonic feedback from galaxies have been performed within the IllustrisTNG cosmological simulations \citep{nelson18,springel18,pillepich18,naiman18,marinacci18}. Recent work by \cite{Garcia:2020kxm} has specifically considered the result of the baryonic feedback in IllustrisTNG that leads to the appearance of cosmological-scale magnetic bubbles, with magnetic field values in excess of $B\gtrsim 10^{-12}$~cG (comoving Gauss), that occupy up to $15\%$ of the simulation volume. \citet{bondarenko21}  studied the effect of these bubbles on searches for the IGMF using \gr\ measurements. This technique is sensitive mainly to the magnetic fields in the voids of the LSS and showed that the presence of such magnetized bubbles has only a minor effect on the \gr\ measurements. However, unlike \gr\  measurements that are sensitive only to the volume-filling IGMF but not to the free electron density $n_{\rm e}$, measurements using the RM technique may be much more influenced by these magnetized bubbles where both $B$ and $n_{\rm e}$ are enhanced. Preliminary estimates from~\citet{Garcia:2020kxm} show that the contribution to RMs from these magnetic bubbles can be comparable to that of the Galactic RM and hence dominate over possible contributions to the RM from the adiabatically compressed primordial magnetic field. 

\begin{figure}
    \centering
    \includegraphics[width=0.48\textwidth]{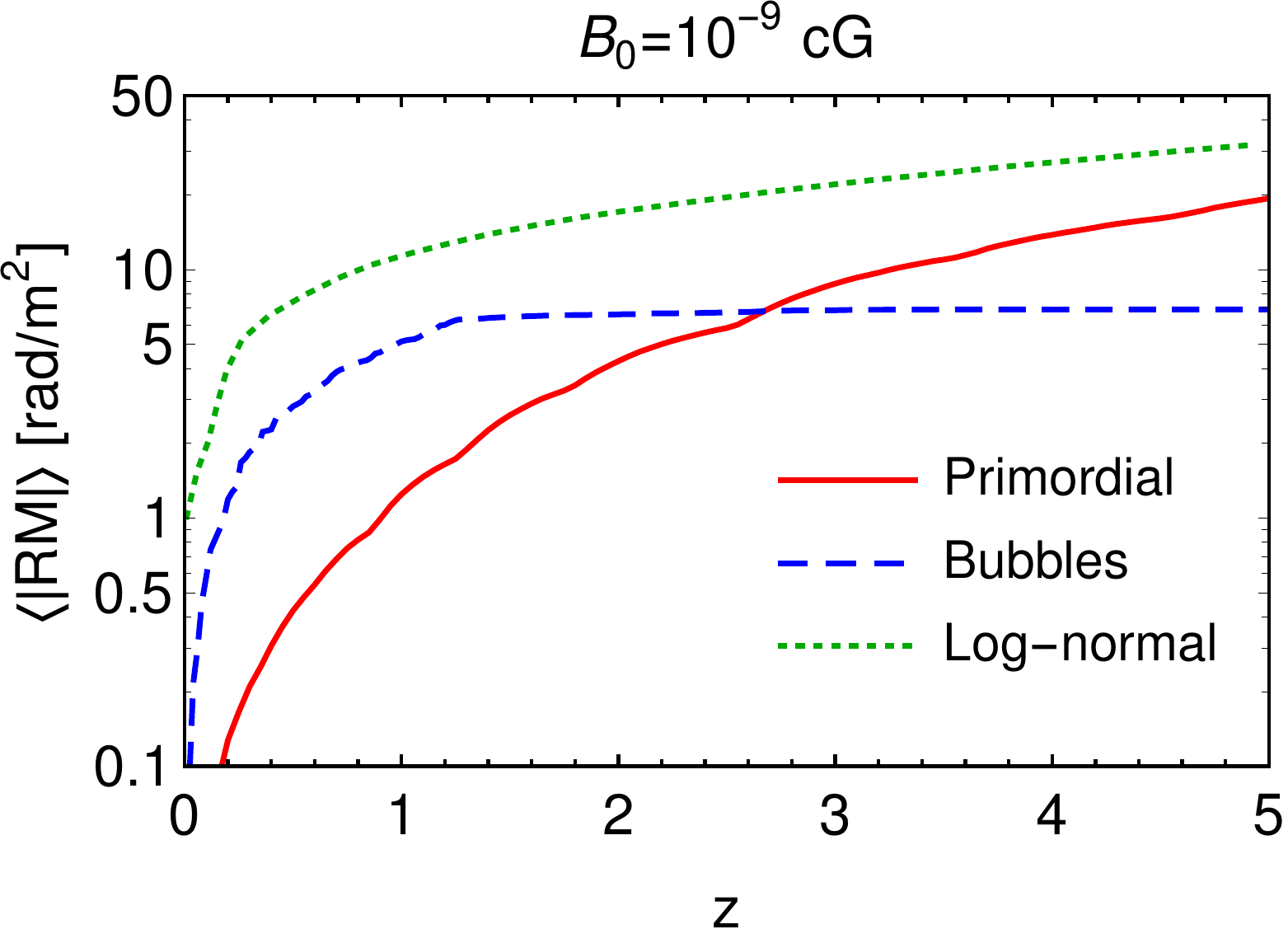}
    \\
    \includegraphics[width=0.48\textwidth]{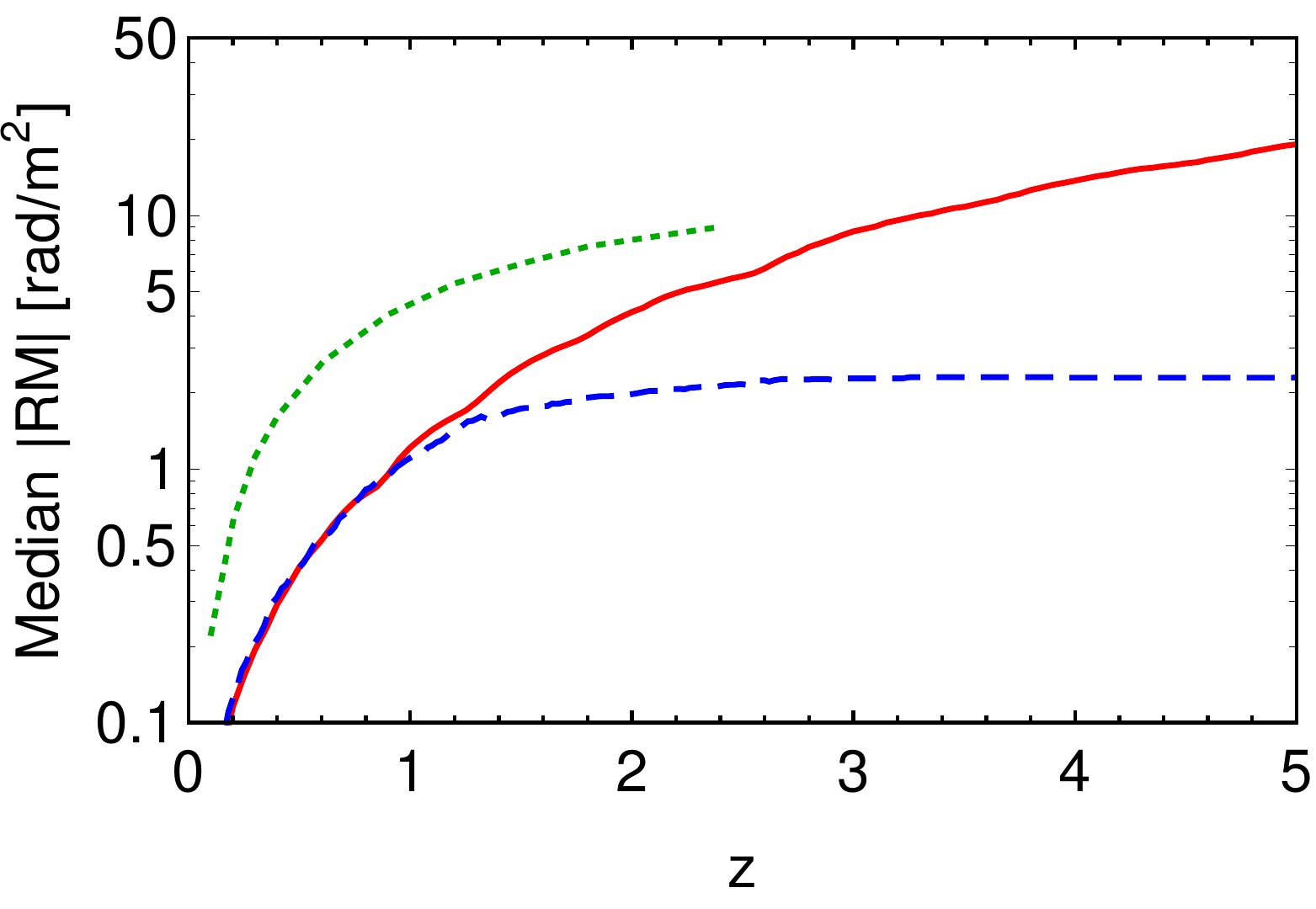}
    \caption[]{Prediction for the mean (upper panel) and median (lower panel) values of $|\text{RM}|$ from the IllustrisTNG simulation as a function of redshift for the homogeneous primordial magnetic field with $B_0 = 10^{-9}$~cG. Red continuous lines show the conservative prediction for the primordial magnetic field, blue dashed lines show the contribution from magnetic bubbles for which we excluded lines of sight with |RM|$>400\text{ rad}/\text{m}^2$ that come from the intersecting galaxies, see text for details. For comparison we also show the prediction from~\cite{Pshirkov:2015tua} (upper panel) and~\cite{Blasi:1999hu} (lower panel), in which RM was estimated based on analytic log-normal distribution for electron number density (green dotted line).}
    \label{fig:prim-vs-bubbles}
\end{figure}

In this work, we make a detailed assessment of the effect of magnetized bubbles around galaxies on the extragalactic RM. We show that the presence of these bubbles can account for a large part of the extragalactic RM at high redshifts $z\gtrsim 2$ and, in fact, that the IllustrisTNG model saturates the current upper limit on extragalactic RM. We also study the consistency of the baryonic feedback model in the IllustrisTNG simulations using RM data from the NVSS. 

The structure of this paper is as follows: in Section~\ref{sec:predictedRM} we describe the IllustrisTNG simulations and the properties of magnetic bubbles; we discuss the separation of the volume-filling component of the IGM from that of magnetic bubbles and extract a redshift-dependent prediction for the RM from the volume-filling magnetic field component and magnetic bubble component, respectively. In Section~\ref{sec:RM-bubbles} we compare our predictions for the RM from magnetic bubbles from the IllustrisTNG simulations to observational measurements from the NVSS survey. In Section~\ref{sec:host-galaxies} we describe a simple analytic model for the RM contribution from host galaxies and compare it to predictions from the IllustrisTNG simulations. In Section~\ref{sec:conclusions} we describe the implications of these results and draw our conclusions. 

Cosmological parameters from \cite{Plank2016A&A...594A..13P} are assumed throughout this work.

\section{Comparing RM from bubbles and primordial magnetic field}
\label{sec:predictedRM}

The IllustrisTNG simulations and the method used here to extract data on rotation measures for random lines of sight are described in detail in our companion paper~\cite{Aramburo-Garcia:2022ywn}. We note that IllustrisTNG is a state-of-the-art gravo-magnetohydrodynamic simulation incorporating a comprehensive model of galaxy formation. In our work, we use the high-resolution TNG100-1 simulation~\citep[hereinafter TNG100 or just TNG; ][]{Nelson2019ComAC...6....2N} with a box size of $\sim (110~\text{cMpc})^3$, which contains $1820^3$ dark matter particles and an equal number of initial gas cells. The initial seed magnetic field in this simulation was chosen to be homogeneous with a magnitude of $10^{-14}$~cG. We divide the simulation volume into magnetic bubbles and primordial magnetic field components using a limiting magnetic field strength of $10^{-12}$~cG as a boundary condition between the two regions (see details in~\citealt{Garcia:2020kxm,Aramburo-Garcia:2022ywn}). The component with $|B|>10^{-12}$~cG is used to make predictions for magnetic bubbles, while the other component we rescale by a factor $B_0/10^{-14}$~cG and used to predict a conservative contribution from the primordial magnetic field with a field strength $B_0$. 

The IllustrisTNG simulation data between redshifts $z=0$ and $5$ is stored in the form of snapshots at 13 redshift points. From each of the snapshots, we extract data for electron number density and magnetic field along 1000 random lines of sight. We found that for some lines of sight intersection of galaxies happened, which resulted in a very large |RM| value. From all the 13000 lines of sight, we exclude four lines of sight that are strong outliers with |RM|$>400\text{ rad}/\text{m}^2$, see more details about outliers in Appendix~\ref{app:RM-distribution}. We create 1000 continuous random lines of sight between redshifts $z=0$ and $z=5$ following the procedure described in~\cite{Aramburo-Garcia:2022ywn}. 

In Figure~\ref{fig:prim-vs-bubbles} we show the predictions for the mean (upper panel) and median (lower panel) absolute RM value, $|\text{RM}|$, for the primordial magnetic field and for magnetic bubbles using $B_0 = 10^{-9}$~cG. The two contributions have different redshift dependence: the RM from the primordial magnetic field grows steadily with redshift, while the prediction from magnetic bubbles saturates around $z\sim 1.5$. This comes from the fact that magnetic bubbles are formed only at small redshifts $z\lesssim 2$~\cite{Garcia:2020kxm}, while the primordial magnetic field exists at all redshifts.

We see that for $B_0 = 10^{-9}$~cG, the contribution from magnetic bubbles dominates the mean |RM| at low redshifts and approximately equal contribution at large redshifts. For the median value, the contribution from magnetic bubbles is more modest, resulting in the strong dominance of the primordial magnetic field contribution at large redshifts.
These differences between the mean and median $|\text{RM}|$ values are due to the RM distribution for bubbles having a long high-RM tail that significantly influences estimates of the mean |RM|. For comparison we also show predictions from~\cite{Pshirkov:2015tua} and~\cite{Blasi:1999hu}. One can see that those previous results show a different redshift dependence to that of our primordial RM. This can be explained by the differences in electron number density distribution between the analytical model of \cite{Pshirkov:2015tua,Blasi:1999hu} and the numerical model of IllustrisTNG, see  \citet{Aramburo-Garcia:2022ywn} for details.

In the following sections, we consider the case that the $B_0$ value is small enough ($B_0 \ll 10^{-10}$) to make its contribution negligible compared to that of magnetic bubbles.

\begin{figure}
    \centering
    \includegraphics[width=\linewidth]{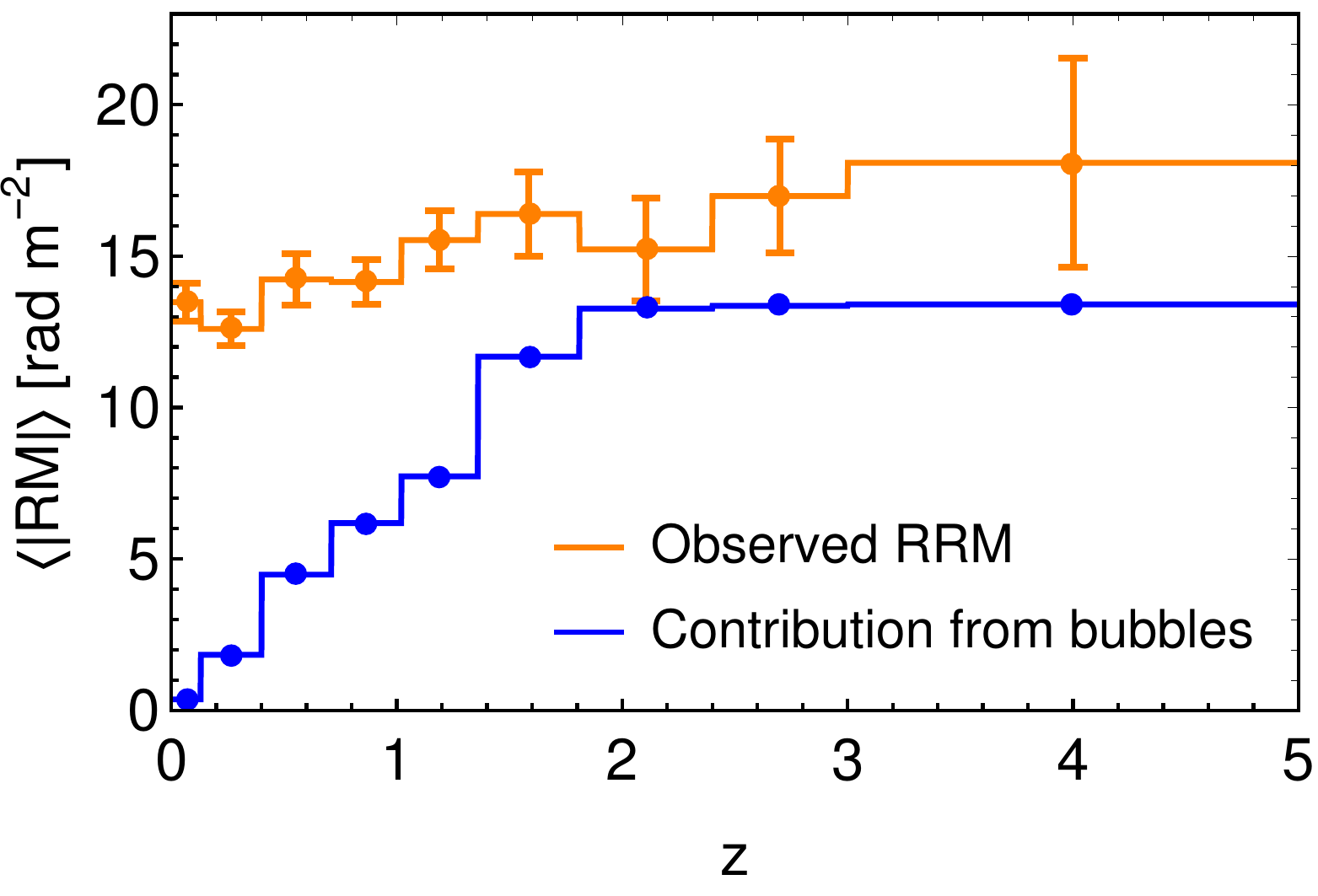}
    \\
    \includegraphics[width=\linewidth]{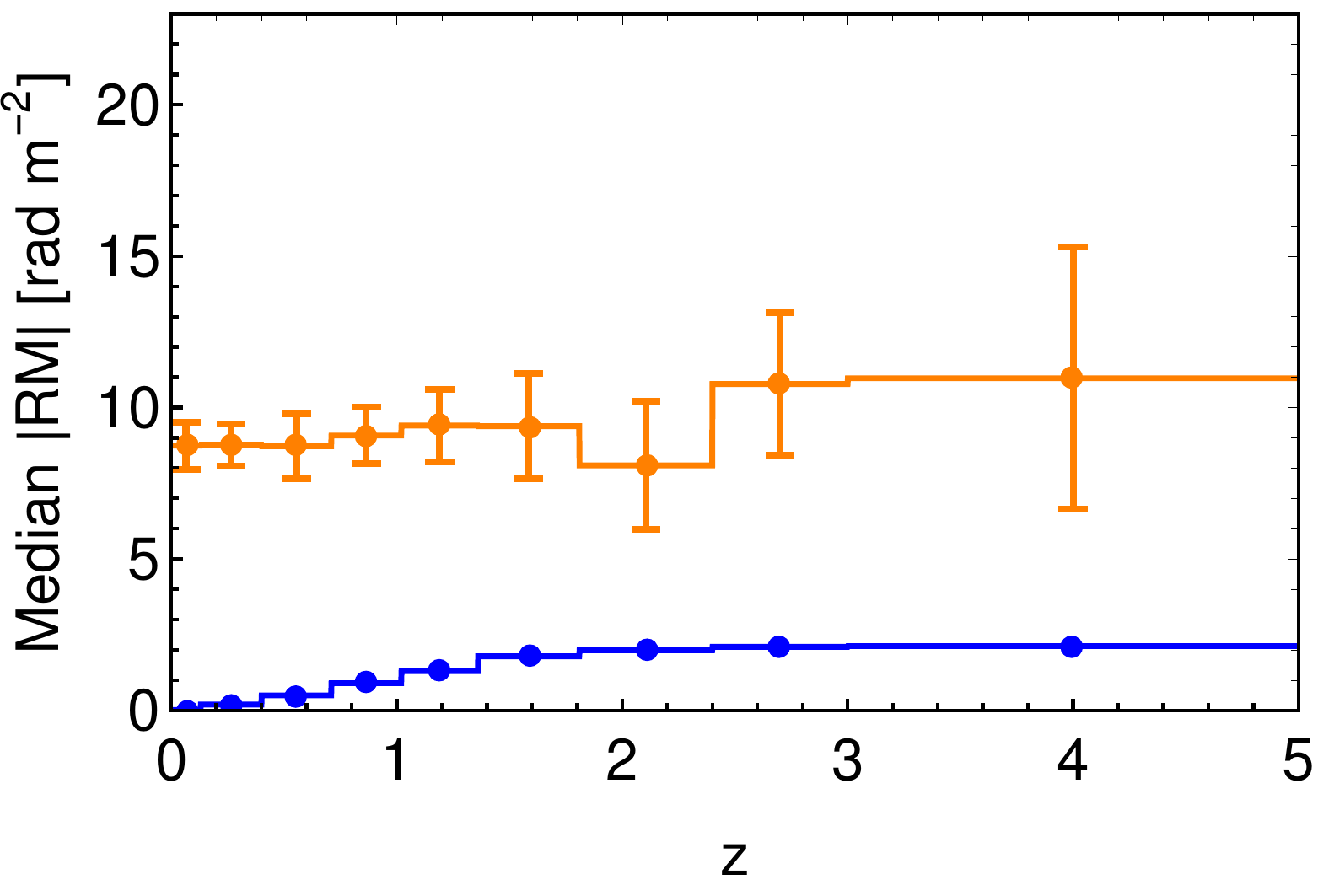}
    \caption{Mean (upper panel) and median (lower panel) observed $|\text{RRM}|$ at different redshifts (orange lines) and contribution from magnetic bubbles in TNG simulation calculated using 1000 random lines of sight (blue lines).}
    \label{fig:AvRMbubbles-vs-AvRRM-RMvoxel-20}
\end{figure}

\section{Comparison between TNG model of magnetic bubbles and observations}
\label{sec:RM-bubbles}

We use observations of 3650 radio sources with Faraday rotation measures, and redshift information cataloged by~\cite{2012arXiv1209.1438H}, where objects close to the Galactic plane ($\ell < 20^\circ$) are removed. 
These data were produced from the NRAO VLA Sky Survey~\citep{Condon1998,Taylor2009} catalog, in which polarization was measured at two close frequencies. This results in a wrapping uncertainty~\citep{Taylor2009}, which means that one cannot distinguish RMs that differ by integer multiples of $\delta\text{RM} = 652.9 \text{ rad}/\text{m}^2$. Therefore, all absolute RM values in the catalog are smaller than $520\text{ rad}/\text{m}^2$, and this is taken into account when we compare them to simulations. We estimate the extragalactic contribution as the residual rotation measure (RRM), which we obtain by subtracting the Galactic RM (GRM) using the model of~\cite{Hutschenreuter2020}.

The comparison between observed RRM and our prediction for magnetic bubbles from the IllustrisTNG simulation for both mean and median values of the RRM is shown in Figure~\ref{fig:AvRMbubbles-vs-AvRRM-RMvoxel-20}, where for the simulated data we include lines of sight that intersect galaxies (opposite to Section~\ref{sec:predictedRM}), as we cannot exclude line of sight with crossing galaxies in experimental data. However, we apply a wrapping correction in order to ensure consistency with observations, see Appendix~\ref{sec:wrapping-correction} for details.
From Figure~\ref{fig:AvRMbubbles-vs-AvRRM-RMvoxel-20}, it can be seen that the prediction of the mean $|\text{RM}|$ from bubbles in the simulation grows quickly with redshift and is almost constant at $z>2$. It is also interesting to notice that the prediction for the mean |RM| from magnetic bubbles at large redshifts almost coincides with the observed data.

For median |RM| values at $z\gtrsim 2$, we see that the contribution from magnetic bubbles is much smaller than the observed RRM, so one might naively conclude that the observed extragalactic RM at large redshifts cannot be explained by magnetic bubbles. However, one should keep in mind that for the observed extragalactic RMs, two systematic factors can increase the observed RM, particularly for small RM values. First, the observed extragalactic RMs have large statistical errors for small RM values. Indeed, almost all data points with extragalactic RMs smaller than $10\text{ rad}/\text{m}^2$ have an associated uncertainty that is of the order of the measured value itself. The second factor is that the procedure for measurement of the Galactic RM depends on the extragalactic sources themselves, which introduces a systematic error into the extragalactic RM, see e.g.~\cite{Oppermann:2014cua}. Both these factors result in a wider extragalactic RM distribution, creating a significant bias in the observed median values of the $|\text{RM}|$.

Consequently, from these results, we conclude that the magnetic bubbles in the IllustrisTNG simulation do not contradict the available observational data. 
If the model from the IllustrisTNG simulation is correct, the magnetic bubbles provide a lower bound for future extragalactic RM measurements at $z\gtrsim 2$, with the characteristic property that the mean value is much larger than the median.

\section{Contribution from host galaxies}
\label{sec:host-galaxies}

An additional contribution to the extragalactic RM potentially comes from the host galaxies and could mask that from magnetic bubbles. In this section, we consider a simple model for host galaxies and argue that their contribution at large redshift should be small.

In general, prediction of the host galaxy contribution in simulations is very tricky, as one should properly choose and correctly model sources of polarized radio emission as similar as possible to those present in the observational sample. In this section, we will discuss a simple qualitative model for host galaxies and compare its predictions with those from the IllustrisTNG simulation.

\subsection{Simple analytic model}
\label{sec:host-model}

The RM from the host galaxy at redshift $z$ is given by
\begin{equation}
    \text{RM} \propto \frac{1}{(1+z)^2} \int n_e B_{\parallel} dL,
    \label{eq:RM-host}
\end{equation}
where the integral is taken along the line of sight in the circumgalactic medium of the host galaxy. Let us consider that electron number density near the galaxy $n_{\rm e}$ behaves like a cosmological average and is proportional to $(1+z)^3$. At the same time, the characteristic size of the region that gives a significant contribution to the RM is constant in comoving coordinates, $L\propto 1/(1+z)$. We see that in this case, the $z$-dependence from $n_{\rm e}$ and $L$ in Eq.~\eqref{eq:RM-host} cancels out, so the redshift dependence of the RM from host galaxies is defined only by the evolution of the magnetic field near the galaxy. Magnetic field evolution in the circumgalactic medium was studied in detail by~\cite{2012MNRAS.422.2152B}: using cosmological MHD simulations and analytic models, it was shown that the magnetic field near galaxies grows quickly at large redshifts, then reaches a maximum at some intermediate redshift and slowly decays after that. We expect that the RM contribution from host galaxies should exhibit similar behavior.
\begin{figure}
    \centering
    \includegraphics[width=\linewidth]{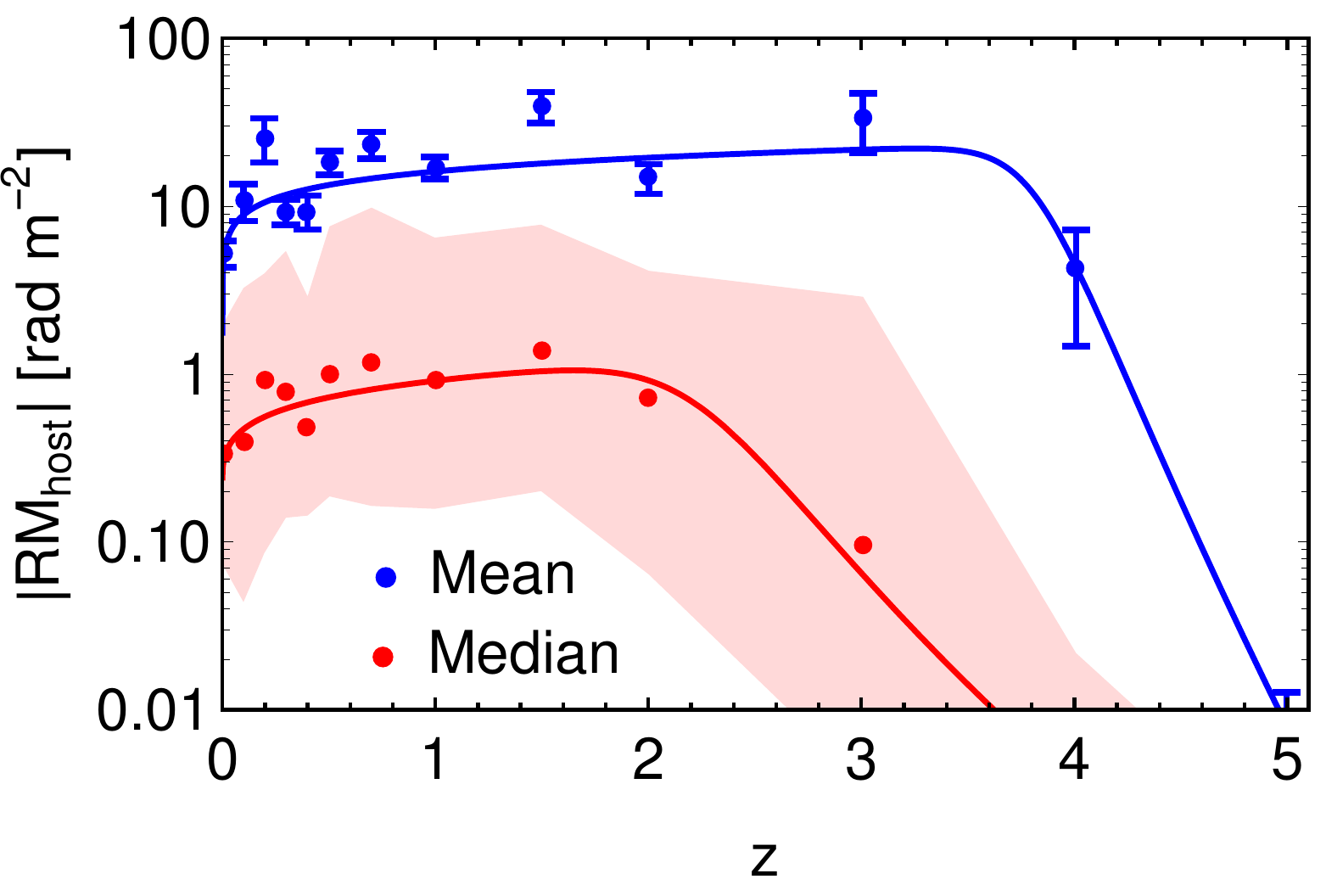}
    \caption{Prediction for the mean (blue points) and median (red points) $|\text{RM}|$ values from host galaxies in the IllustrisTNG simulation. With error bars for the mean, we show an estimate of the standard errors, while the shaded red region shows a central region in which $50\%$ of data lies. By continuous lines of corresponding colors we show simple broken power-law fits to the data.}
    \label{fig:RM-host}
\end{figure}

\subsection{Host galaxies in IllustrisTNG}

To model host galaxies in the IllustrisTNG simulation, we assume that the main sources of observed RM are the radio lobes at the end of AGN jets. We assume that two radio lobes are located symmetrically around an AGN and that we cannot resolve these two radio lobes in the observational data. We also assume that both radio lobes have the same intensity of polarized emission so that the observed RM is an average of their individual RMs.

For each given redshift in the simulations, we choose $\sim100$ random galaxies that contain supermassive black holes (the minimal mass of the supermassive black hole in IllustrisTNG is $10^6M_{\odot}$, and it is placed in the center of each dark matter halo when it reaches a virial mass of $6\cdot 10^{10}M_{\odot}$). For these galaxies, we generate two symmetric radio lobes pointing in a random direction from the galaxy center with an isotropic distribution and a randomly selected distance between the two radio lobes in the range from 50 to 300\,kpc. These distances are chosen according to the experimentally measured distribution from~\cite{2020MNRAS.499...68T}. For each pair of radio lobes, we generate six lines of sight in random directions and use the first 1\,Mpc along these lines of sight to define the contribution from the host galaxy. 

Using these data we calculate the mean and median |RM| from host galaxies, the result of which is shown in Figure~\ref{fig:RM-host}. The continuous lines show best fits to the model
\begin{equation}
    |\text{RM}|(z) = \frac{a + b z^c}{1 + (z/d)^e},
\end{equation}
with best-fit parameters $a=1.65(0.242)\text{ rad}/\text{m}^2$, $b=14.4(0.670)\text{ rad}/\text{m}^2$, $c=0.300(0.468)$, $d=3.81(2.26)$, $e = 29.8(10.6)$ for mean (median). Qualitatively, the RM from host galaxies changes with redshift according to the simple analytic model from Section~\ref{sec:host-model}: it decays at large redshifts and has a maximum at intermediate redshifts of $z\sim 3$ for the mean and $z\sim 1.5$ for the median. Comparing the contribution from host galaxies with Figure~\ref{fig:prim-vs-bubbles} we conclude that within the IllustrisTNG model the RM of high redshift sources ($z>3$) is dominated by the contribution from magnetic bubbles along the line-of-sight, rather than by the host galaxy RM, for
both the mean and median absolute RM values.

\section{Discussion and Conclusions}
\label{sec:conclusions}

In this work, we have considered the effect of magnetized bubbles around galaxies driven by baryonic feedback processes on the extragalactic Rotation Measure. We have used the IllustrisTNG simulation to separate the contributions of the volume-filling intergalactic magnetic fields and the contribution of magnetized outflows from galaxies to the RM integral. We have demonstrated that the IllustrisTNG model of such magnetized bubbles predicts that the extragalactic RM at $z>2$ almost saturates current estimates of the mean residual RM from the NVSS, see Figure~\ref{fig:AvRMbubbles-vs-AvRRM-RMvoxel-20}. The contribution of magnetized bubbles to the extragalactic RM at $z>2$ (including wrapping correction, see Appendix~\ref{sec:wrapping-correction}) has a value of $\langle |{\rm RM}| \rangle \simeq 13$~rad/m$^2$, which is very close to the mean residual RM of $16$~rad/m$^2$ found from NVSS data in this redshift range when accounting for the Galactic RM model of \cite{Hutschenreuter2020}. Without the survey-dependent wrapping correction the prediction for the mean absolute RM from magnetic bubble is $\langle |{\rm RM}| \rangle \simeq 7$~rad/m$^2$, where rare lines of sight with |RM|$>400\text{ rad}/\text{m}^2$ that came from intersecting galaxies were excluded (see Section~\ref{sec:predictedRM} for details). 

While our work suggests that there are two main contributions in the IGM: (i) from magnetic bubbles and (ii) from the volume-filling magnetic field, the results found here indicate that the contributions from these two components have different redshift dependencies: the volume-filling magnetic field exists at all redshifts, and its contribution constantly grows with $z$, while magnetic bubbles are formed at later times, mostly below $z \approx 2$, and so at larger redshifts their contribution is fixed. This should allow one to distinguish the separate contributions in future observations.

We also consider a simple analytic model for the RM contribution from host galaxies and confirm it using data from the IllustrisTNG simulation. We show that the contribution from host galaxies to the mean and median |RM| values quickly decreases at large redshifts. 
This provides a possibility to isolate the contribution of magnetic bubbles along the line of sight into the overall extragalactic RM. This can be done through a comparison of the RM of high-redshift sources ($z\gtrsim 2...3$) that of the lower redshift sources. If the main source of the extragalactic RM is the magnetic field around the source host galaxies, high-redshift sources should have systematically lower RM. A caveat of this approach may be the cosmological evolution of the source population, which is not considered in our simple source model (radio lobes around the host galaxy).

While the predicted mean absolute RM from the IllustrisTNG simulations found here is compatible with observational measurements from the NVSS survey, we note that the wrapping correction implemented in this work may represent a systematic uncertainty in this result that artificially lowers the predicted and measured RM. Next-generation polarisation surveys such as those expected from the SKA telescope and its precursors will not be subject to this same wrapping uncertainty in their RM measurements due to the broadband nature of their measurements. Given the closeness of the current IllustrisTNG model predictions to the observed residual RM estimates derived from current data suggested that the IllustrisTNG model of baryonic feedback will be falsifiable with the improvement of RM measurements from these new surveys. Furthermore, compared to the NVSS data considered in this work, the SKA will provide an RM grid containing several orders of magnitude more sources than the $\sim 4\times 10^3$ source sample considered here. 
A denser RM grid provided by the new surveys and better precision stemming from broadband rather than a two-frequency sampling of the polarised signal will also improve the knowledge of the Galactic component of the RM. This will result in smaller systematic uncertainty of the RRM (specifically for the median of the absolute value, which is possibly dominated by the systematic uncertainty). 
If the RRM level found with the SKA data is lower than the current estimates derived from NVSS, the IllustrisTNG model will be in tension with the data. This suggests that the IllustrisTNG baryonic feedback model is falsifiable through the RM measurements.

\section*{Acknowledgements}

KB is partly funded by the INFN PD51 INDARK grant.
AB is supported by the European Research Council (ERC) Advanced Grant ``NuBSM'' (694896). AMS gratefully acknowledges support from the UK Alan Turing Institute under grant reference EP/V030302/1. AS is supported by the Kavli Institute for Cosmological Physics at the University of Chicago through an endowment from the Kavli Foundation and its founder Fred Kavli. This work has been supported by the Fermi Research Alliance, LLC under Contract No. DE-AC02-07CH11359 with the U.S. Department of Energy, Office of High Energy Physics.


\section*{Data Availability}

The data underlying this article is available on reasonable request.

\bibliographystyle{mnras}
\bibliography{refs.bib}

\appendix

\section{High-RM distribution tail in simulations}
\label{app:RM-distribution}

\begin{figure}
    \centering
    \includegraphics[width=\linewidth]{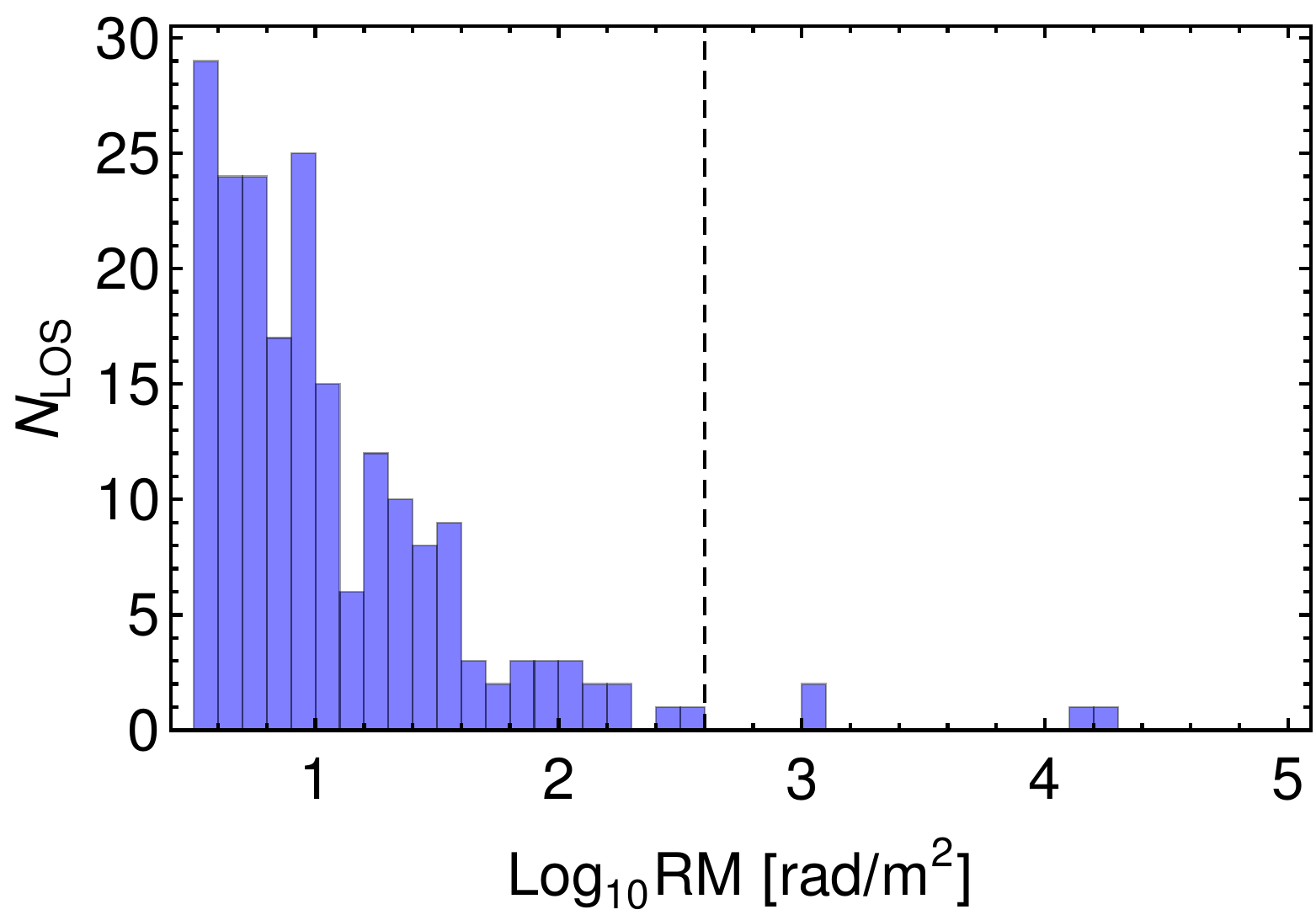}
    \caption{High-RM tail of the distribution of |RM| for 13000 lines of sight extracted from the IllustrisTNG simulation box between redshifts $0$ and $5$. Black dashed line corresponds to |RM|$=400\text{ rad}/\text{m}^2$.}
    \label{fig:High-RM-tail}
\end{figure}

In Figure~\ref{fig:High-RM-tail} we show the high-RM tail for the distribution of the total RM for 13000 lines of sight extracted from IllustrisTNG simulation. We see that in this data there are 4 outliers with |RM| larger than $1000\text{ rad}/\text{m}^2$. We checked that all of them correspond to the intervening galaxy along the line of sight. We exclude them for prediction shown in Figure~\ref{fig:prim-vs-bubbles} as these outliers strongly influence average |RM|, but can be easily excluded from the experimental data using condition |RM|$<400\text{ rad}/\text{m}^2$ or by techniques discussed in e.g.~\cite{Farnes:2014oaa}. To study more accurately the influence of these outliers, more simulation data is needed, which is computationally hard. We will leave this study for future studies.

\section{Observational data in bins}
\label{app:errors}

\begin{table*}
\centering
\begin{tabular}{|l|r|r|r|r|r|r|r|r|r|}
\hline
Bin number & 1 & 2 & 3 & 4 & 5 & 6 & 7 & 8 & 9 \\
\hline
Upper bound for $z$ & $0.13$ & $0.40$  & $0.71$  & $1.02$  & $1.36$  & $1.81$ & $2.4$ & $3$ & $5$   \\ \hline
Object number & $564$  & $781$   & $500$   & $441$   & $426$   & $437$   & $322$ & $129$ & $49$   \\ \hline
$\langle |\text{RRM}|\rangle$, rad$/$m$^2$ &
$13.48$ & $12.60$ & $14.23$ & $14.15$ & $15.53$ & $16.39$ & $15.22$ & $16.98$ & $18.08$ \\ \hline
$\Delta\langle |\text{RRM}|\rangle$, rad$/$m$^2$ & $0.62$ & $0.56$ & $0.85$ & $0.75$ & $0.96$ & $1.39$ & $1.69$ & $1.87$& $3.45$ \\ \hline
$\text{Med}\,|\text{RRM}|$, rad$/$m$^2$ &
$8.74$ & $8.77$ & $8.72$ & $9.07$ & $9.41$ & $9.38$ & $8.09$ & $10.77$ &
$10.97$ \\ \hline
$\Delta\text{Med}\,|\text{RRM}|$, rad$/$m$^2$ &
$0.78$ & $0.70$ & $1.07$ & $0.94$ & $1.20$ & $1.74$ & $2.12$& $2.34$ & $4.32$ \\ \hline
\end{tabular}
\caption{Summary of the observational data in bins used in this work. The first row shows the bin number, the second row shows upper bound on each redshift bin (the lower bound of the first bin is $z=0$). In the third row we show the number of observed objects, also we show the mean $|\text{RM}|$ and $|\text{RRM}|$ and their errors in each bin.}
\label{tab:zbins-new}
\end{table*}

In this work, we bin observational data in 9 bins with an approximately equal number of objects at small redshifts. In each bin, we calculate the mean and median values of $|\text{RM}|$ and estimate their statistical errors. For mean, we calculate the standard error $\Delta x$ in each bin as
\begin{equation}
    \Delta \langle x \rangle = \sqrt{\frac{\langle(x_i - \langle x \rangle)^2\rangle}{n}},
\end{equation}
where $x_i$ are $|\text{RRM}|$ values in each bin, $\langle x \rangle$ is their mean value, and $n$ is a number of objects in the bin. In the same notation, the error of the median is estimated as~\citep{williams_2001}
\begin{equation}
    \Delta \text{Med}(x) = \sqrt{\frac{\pi}{2} \frac{\langle(x_i - \langle x \rangle)^2\rangle}{n}} \approx 1.253 \cdot \Delta \langle x \rangle.
\end{equation}
We summarize our results in Table~\ref{tab:zbins-new}.

\section{Wrapping correction}
\label{sec:wrapping-correction}

The data for rotation measure in the catalog that we use in this work can be subject to a wrapping uncertainty with step $\delta \text{RM} = 652.9\text{ rad}/\text{m}^2$. In our theoretical prediction, lines of sight sometimes appear with RMs of order $\mathcal{O}(1000)\text{ rad}/\text{m}^2$, so it is important to make a wrapping correction similar to experimental data if we want to compare them. Of course, we do not have depolarization data, but based on the description of the procedure, we emulate it in the following way:
\begin{enumerate}
    \item If the absolute value of the RM is smaller than $520\text{ rad}/\text{m}^2$ we do not change it.
    \item If $|\text{RM}|>520\text{ rad}/\text{m}^2$ we take the value $\text{RM} + N \delta\text{RM}$, where $N$ is such integer number (positive or negative) such that the resulting RM has the smallest absolute value.
\end{enumerate}
%

\bsp	
\label{lastpage}
\end{document}